\documentclass[letterpaper,rmp,twocolumn,amsmath,amssymb,groupedaddress,floatfix]{revtex4}
\RequirePackage[sort&compress]{natbib} %References in order as cited

\usepackage[utf8]{inputenc}
\usepackage[T1]{fontenc}
\usepackage{ae}
\usepackage{aecompl}

\usepackage[a]{esvect}

\usepackage{graphicx}% Include figure files
\usepackage{dcolumn}% Align table columns on decimal point
\usepackage{bm}% bold math

\usepackage{pgfplots}

\definecolor{darkgray}{rgb}{0.5,0.5,0.5}
\definecolor{darkred}{rgb}{0.89,0.10,0.11}
\definecolor{darkblue}{rgb}{0.12,0.39,0.62}
\usepackage{url}
\usepackage{enumitem}
\usepackage{tikz}
\usetikzlibrary{calc}

\usepackage[breaklinks=true,colorlinks=true,citecolor=black,linkcolor=black,menucolor=black,urlcolor=darkblue,pdfborder={1 0 0}]{hyperref}
\hypersetup{pdftitle={Stock portfolio structure of individual investors infers future trading behavior},pdfauthor={L. Bohlin and M. Rosvall (2014)}}
\bibpunct{(}{)}{,}{n}{,}{,}

\usepackage[font=sf,size=footnotesize,justification=RaggedRight,singlelinecheck=false]{caption}
\usepackage[sf,bf,small]{titlesec}
\titlespacing*{\section}{0pt}{1em}{0em}
\titlespacing*{\subsection}{0pt}{1em}{0em}
\titlespacing*{\subsubsection}{0pt}{1em}{0em}

\renewcommand{\thefigure}{\arabic{figure}}

\renewcommand\thesection{\arabic{section}}
\renewcommand\thesubsection{\thesection.\arabic{subsection}}

\usepackage{booktabs}

\newcommand{\mybottomrule}{\specialrule{0.1em}{0em}{0em}}

\makeatletter
\renewcommand{\p@section}{\arabic{section}\expandafter\@gobble}
\renewcommand{\p@subsection}{\thesection.\arabic{subsection}\expandafter\@gobble}
\renewcommand{\p@subsubsection}{\thesubsection.\arabic{subsubsection}\expandafter\@gobble}
\makeatother

\titleformat*{\section}{\large\bfseries\sffamily}
\titleformat*{\subsection}{\normalsize\bfseries\sffamily}
\titleformat*{\subsubsection}{\normalsize\sffamily}

\everymath=\expandafter{\the\everymath\displaystyle}

\begin{document}
\makeatletter
\renewcommand\@biblabel[1]{#1.}
\makeatother

%\preprint{APS/123-QED}

%%%%%%%%%%%%%%%%%%%%%%%%%%%%%%%%%%%%%%%%%%%%%%%%%%%%%%%%%%%%%%%%%%%%%%%%%%%%
%%%%%%%%%%%%%%%%%%%%%%%%%%%%%%%%%%%%%%%%%%%%%%%%%%%%%%%%%%%%%%%%%%%%%%%%%%%%
%%%%%%%%%%%%%%%%%%%%%%%%%%%%%%%%%%%%%%%%%%%%%%%%%%%%%%%%%%%%%%%%%%%%%%%%%%%%
%%%%%%%%%%%%%%%%%%%%%%%%%%%%%%%%%%%%%%%%%%%%%%%%%%%%%%%%%%%%%%%%%%%%%%%%%%%%
\title{Stock portfolio structure of individual investors infers future trading behavior}% Force line breaks with \\

\author{Ludvig Bohlin}
\email[]{ludvig.bohlin@physics.umu.se}
\thanks{Corresponding author}
\author{Martin Rosvall}%
\email[]{martin.rosvall@physics.umu.se}
\affiliation{%
Integrated Science Lab, Ume\r{a} University, Sweden\\
}%

\date{July 28, 2014}% It is always \rightarrowday, today,
             %  but any date may be explicitly specified

% Please keep the abstract between 250 and 300 words
\begin{abstract}
Although the understanding of and motivation behind individual trading behavior is an important puzzle in finance, little is known about the connection between an investor's portfolio structure and her trading behavior in practice. In this paper, we investigate the relation between what stocks investors hold, and what stocks they buy, and show that investors with similar portfolio structures to a great extent trade in a similar way. With data from the central register of shareholdings in Sweden, we model the market in a similarity network, by considering investors as nodes, connected with links representing portfolio similarity. From the network, we find investor groups that not only identify different investment strategies, but also represent individual investors trading in a similar way. These findings suggest that the stock portfolios of investors hold meaningful information, which could be used to earn a better understanding of stock market dynamics.

\end{abstract}

\pacs{Valid PACS appear here}% PACS, the Physics and Astronomy
                             % Classification Scheme.
%\keywords{Suggested keywords}%Use showkeys class option if keyword
                              %display desired
\maketitle

% Please keep the Author Summary between 150 and 200 words
% Use first person. PLoS ONE authors please skip this step. 
% Author Summary not valid for PLoS ONE submissions.   
%\section*{Author Summary}

\section*{Introduction}
Stock market trading provides opportunities at the cost of risk. For investors, the ultimate trading goal is to make as much money as possible by acting in such a way that the highest possible profit is realized at minimal risk. Yet how to trade optimally is far from obvious, and many factors influence trading behavior. The traditional approach to represent market trading has been a model perspective, assuming that investors act as rational and identical agents \cite{hommes2006heterogeneous, barberis2003survey}. However, this assumption has been challenged by empirical evidence, which suggests that also other elements are present in financial markets \cite{barber2011behavior}. For example, researchers have suggested models in which the economic decisions of investors also consider the effects of social, cognitive, and emotional factors. These factors and their influence on trading are often studied with data external to the actual trading process and include, for example, proximity \cite{bodnaruk2009proximity}, social media interactions \cite{bollen2011twitter} and web engine search queries \cite{preis2013quantifying, moat2013quantifying}, In this work, we instead focus on data more directly relevant to the trading by relating trading patterns and behavioral biases to stock portfolio structure. With real financial data on individual investors from the Swedish stock market, we study the connection between what stocks investors hold, and what stocks they buy.

Trading by individual investors and related behavioral biases have been studied by, for example, local bias \cite{seasholes2010individual}, overconfidence \cite{odean1998volume}, sensation seeking \cite{grinblatt2009sensation} and the disposition effect \cite{shefrin1985disposition}. Studies show that investors base trading not only on rationale information, but are also affected by factors connected to both personal characteristics and associated external conditions.  Various individual biases give rise to a trading heterogeneity among investors, or among groups of investors. The presence of such investor groups with similar trading behavior can be explained by homophily, i.e., the tendency of individuals to behave and bond with others who are similar. Investors potentially trade more similarly if they share certain properties, including, for example, age \cite{korniotis2011older}, gender \cite{barber2001boys} and familiarity \cite{massa2006hedging}. However, classifying individual investors into such distinct categories is not straightforward. The classification is in some cases motivated by theoretical considerations, and in other cases by observed patterns in the data. Categories include, for example, fundamentalists and chartists \cite{frankelfroot}, and informed and uninformed traders \cite{da2005informed}. Other examples of investor categorizations in financial data are derived from trading correlations \cite{iori2007trading}, direct stock trading data \cite{wang2011characteristics}, and network approaches \cite{tumminello2012identification, jiang2010complex}.

Although the understanding of and motivation behind individual trading behavior is an important puzzle in finance, little is known about the connection between investors' portfolio structure and their trading behavior in practice. Studies have found that many individual investors tend to hold poorly diversified portfolios and instead concentrate investments in only a few stocks \cite{calvet2006down}. The difficulty of searching through all available stocks also makes it more likely for individual investors to invest in stocks that attract their attention \cite{barber2008all}, and these attention-grabbing stocks are typically the ones that investors already hold in their portfolios. This bias indicates that portfolio structure and trading decisions are naturally connected.

In this paper, we explore the connection between portfolio structure and trading behavior in individual investors. With detailed data on stock portfolios from the central register of shareholdings in Sweden, we study the relation between stock portfolio similarity and trading similarity. Unlike most previous research, we do not analyze the direct trading, but instead focus on long-term trading behavior by looking at changes in share portfolios over time. We aim to examine two main questions: (1) How do investors in the stock market structure their portfolios? And (2) Can we learn about trading behavior by looking at the investors' portfolio structure?

To answer the questions of individual trading behavior we take three steps: First, we investigate how individual investors hold stocks, and how they trade. Second, we divide investors into groups based on portfolio similarity. This division is done with a network approach, where individual investors are considered to be nodes, and links between investors are constructed according to stock portfolio similarities. To group similar investors, we analyze the network with the community detection algorithm Infomap \cite{rosvall2008maps}. Third, we analyze the derived groups to investigate the relationship between portfolio structure and trading behavior. This analysis is done by comparing investor trading within groups to investor trading outside the group. In the following section we present the methods and the results, and, in short, we find that the portfolio structure of individual investors holds information on trading behavior, and that investors with similar portfolios, to a great extent, trade in a similar way.

%The authors emphasize the importance of studying portfolio data to analyse market dynamics. These insights based on traditional data sources were recently supported and augmented by large scale analyses on human (online) interaction. For example Bollen, Mao, and Zeng (2011) provided evidence that mood on Twitter affects the stock market. Preis, Moat, and Stanley (2013) demonstrated that search volume collected by Google can be used to predict subsequent stock market moves. Moat, Curme, Avakian, Kenett, Stanley, and Preis (2013) discovered support for this finding by drawing on records on how often Wikipedia pages in financial topics were viewed. This recent series of papers demonstrates the importance of behavioural trends for stock markets in addition to traditional data sources

% You may title this section "Methods" or "Models". 
% "Models" is not a valid title for PLoS ONE authors. However, PLoS ONE
% authors may use "Analysis" 
\section*{Methods}
\subsection*{Data from the central register of Swedish \\ shareholdings}
We examined more than 100,000 individual investors who were actively trading in the Swedish stock market from 2009 to 2011. The investors and their stock portfolios were extracted from a dataset with around two million investors. The dataset stems from the central register of shareholdings in Sweden, and covers basically all investors and their holdings in every publicly traded company in Sweden. The dataset was provided by Euroclear Sweden AB, and permission to use the data was given under a special agreement. Data are presented in half-year share register reports between June 30, 2009, and December 30,  2011, with detailed ownership information of investors in each registered company. The reports also included the companies' total share amount and their corresponding stock ISIN (International Securities Identification Number). Additional data, obtained from the Swedish Central Statistics Office (SCB), provided share prices for companies listed on the Stockholm stock exchange. Those data specify share prices at stock exchange closing time, i.e., the price of the latest sold share on the last trading day. If price data are lacking, bid price and then ask price were used instead. In total, the data contain share prices for around 500 listed stocks. The full dataset makes it possible to extract the detailed portfolio of an investor in the Swedish stock market.  We have made anonymized and reduced data available online as detailed in Ref.~\cite{Data_portfolio}. Below, we explain the dataset in more detail, and the restrictions we set on the data in the analysis. 

Investors are reported either as legal persons, e.g.~corporations and funds, or natural persons, i.e., individual investors. Since we are interested in the trading behavior of individual investors, we considered investors that actually changed their portfolio, and focused on the holdings that investors can manage themselves, namely, direct holdings. Direct holdings are registered in the investor's name, as opposed to nominee holdings, which are registered and managed by an equity manager on behalf of the investor. The direct holdings of all investors are presented in each half-year report, with detailed information about registration type, share amount, and the equity ISIN in which the shares are held. This information makes it possible to find share changes in the portfolios between reports, provided that investors have a traceable identification number. Investors who lack a Swedish identification can not be reliably tracked in the data over time, and we therefore excluded these investors in the analysis.

To reduce the effects of noise in the data, some conditions for the included stocks were established. First, we only considered stocks of companies that existed for the entire time period. We therefore excluded stocks that were introduced or removed from the market during the time period for any reason. This exclusion was done to enable comparisons between two share reports without changes in the company domain. Furthermore, we also required that the total share amount of a stock must not have changed more than five percent during the time period. This condition was set because larger changes make it hard to distinguish actual active trades of investors from more passive changes in the portfolios directly related to a share amount change, as, for example, in the case of stock splitting. As a consequence of the share amount change criteria, we excluded, for example, the companies H\&M and Swedish Match from the analysis. Finally, only listed stocks were considered in the analysis, since these stocks are publicly traded and it is possible to find an explicit price for them. It is also worth noticing the distinction between stocks designated A and B on the Swedish market. A company can be associated with more than one stock, because A and B stocks, and other potential stock classes in a company, must have different ISIN codes. We considered these different stock classes as separate stocks, because classes with less voting power usually are more liquid, and therefore give rise to different trading than the ones with superior voting rights. 

In summary, we examined investors on the Swedish stock market who are natural persons, traceable over time, active in trading and primarily registered as shareholders. This selection means that we, for example, excluded investors registered as legal persons and secondary ownership through funds. We required company stocks to be listed and stable, in the sense that they must exist for the entire time period and have a share amount that do not change too drastically over time. With this noise reduction and data cleaning, we were left with 100,161 investors holding capital in 209 different stocks. 

\subsection*{Portfolio vectors and trading vectors}
We represent investor holdings in normalized portfolio vectors and consider the stock portfolio of an investor as a vector $p$, where $p_i$ represents the investor's proportion of capital  invested in stock $i$. As an example, we can look at an investor with portfolio vector $p$ in a market with four stocks. If the investor holds shares of total value 20 in stock 1, and shares of total value 80 in stock 3, the portfolio vector can be expressed as $p = (0.2, 0, 0.8, 0)$. Note that the value at a specific index represents the relative amount invested in corresponding stock, rounded to the nearest hundredth in the analysis for simplicity. 

The portfolio vector representation is used to unify investor trading, even though the shareholding data reports do not provide direct trading information. However, the reports do specify detailed half-year snapshots of the investors' portfolios, and these snapshots make it possible to track changes in the portfolios over time. Analogously to portfolio vectors, we therefore construct trading vectors for investors based on the sum of all changes between two dates. In these trading vectors, we considered stocks in which investors bought shares during the time period. We only examined purchases, because correlations between portfolio structure and sold stocks follow trivially since investors can buy any stock but only sell stocks they already hold. To compute the trading vector, $p_T$, we therefore extracted the positive changes in the portfolio, $p_T = p(t_2)-p(t_1)$, between reports from June 30, 2009, and December 30, 2011, with stock prices from the first date. In this way, all elements in the trading vectors become positive. To examine the connection between portfolios and trading, we computed a similarity value between investors' portfolio and trading vectors, based on cosine similarity \cite{strehl2000impact}. Accordingly, the similarity of vectors $x$ and $y$ is given by the normalized dot product
\begin{equation*}
sim(\mathbf x,\mathbf y) = \frac{\langle \mathbf x,\mathbf y \rangle}{||\mathbf x||\cdot ||\mathbf y||}.
\label{CosSim}
\end{equation*} 
We use this similarity measure for the portfolio and trading vectors because it is simple and well-suited for analyzing investment structure, The similarity value is bounded between 0 and 1, since all portfolio and trading vector elements are non-negative. 

\subsection*{Identifying groups of investors with similar portfolio structure}
\begin{figure}[!htbp]
\centering
\includegraphics[scale=0.37]{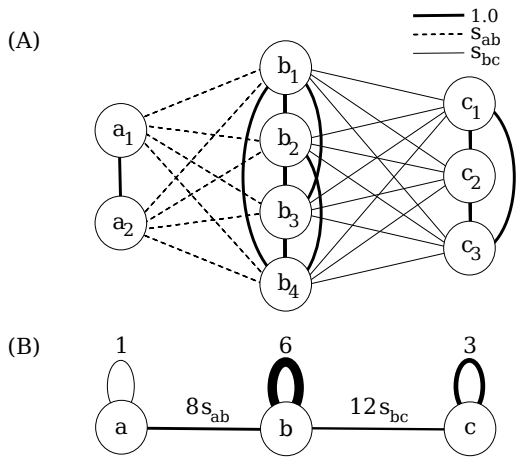}
\caption{\textbf{Network reduction technique.} Example of similarity network reduction from single investor level to portfolio level. (A) Network at individual level with nine investors and three different portfolio structures $a$, $b$ and $c$. Investors are connected with links of value 1, $s_{ab}$ and $s_{bc}$. 
(B) Reduced network at portfolio level, with the three portfolio structures as nodes. Self-links of portfolio nodes represent links between investors with equal portfolio structure. The weight $w_i$ for the self-link of portfolio $i$, with $n_i$ investors, is calculated according to $w_i = n_i(n_i-1)/2$. The link weight $w_{km}$ between portfolios $k$ and $m$, with $n_k$ and $n_m$ investors, respectively, is calculated according to $w_{km} = w_{mk} = n_{k}n_{m}s_{km}$, where $s_{km}$ is the link weight between two investors in each portfolio structure. \label{fig:reduced}}
\end{figure}

Since single investors hold sparse portfolios and trade to a small extent, we need to categorize similar investors in groups, and examine the overall trading behavior of each group. However, dividing investors into groups with similar portfolios is not a straightforward task, since the number of investors is large and it is difficult to distinguish groups without making assumptions and subjective divisions. One possibility would be to simply group investors with the most similar portfolios, but that approach causes problem on where to separate the groups, and we run the risk of losing important structural information. So we require a method of dividing investors into groups that accounts for both portfolio similarity and the structural information of the system. These premises can be fulfilled with clustering tools from network theory, and we therefore take a network approach to analyze the data. Network theory has received increasing interest in finance research recent years, thanks to its ability to model the organization and structure of large complex systems \cite{allen2009networks}. Network approaches aimed to find structures in finance data include, for example, bank-liability networks \cite{boss2004network}, stock correlation networks \cite{huang2009network} and trading networks \cite{jiang2010complex}. In our network approach, we model the data as a network with investors as nodes connected by links according to portfolio similarity. Optimally, we create links between nodes according to causal connections between investors, but such relationships are difficult to obtain, and it is not even clear what they would be. Instead, we use portfolio similarity as a representation of relationships, and connect investor nodes with weighted and undirected links, according to the similarity value of the investors' portfolio vectors. This representation creates a network of investors with links based on portfolio similarity, and we refer to this network as a \textit{similarity network}. To account for the similarity values with the most information and also make the analysis more efficient, we only consider links with values greater than or equal to 0.9 in the similarity network. In the similarity network, investors with at least one common holding will have a similarity value and accordingly be connected by a link. This means that the total number of links in the network can be large for a single investor. To reduce this complexity and make the analysis procedure feasible and more effective, we capitalize on the fact that many investors have identical portfolio vectors and use this to create a reduced network. Investors with equal portfolio structures will share the same links to other investors, which results in a large amount of redundant information. To remove this redundancy and reduce network size, we therefore represent every portfolio structure as a node, instead of having one node for each single investor. This approach reduces both the number of nodes and the number of links in the network. The resulting reduced network becomes an aggregated version of the original network, where links between investors with identical portfolio structure are represented by a self-link. An example of the reduction procedure can be seen in Figure \ref{fig:reduced}. 

The goal of constructing the similarity network is to group investors with similar portfolio structure, albeit not necessarily exactly the same portfolio. To identify candidates for such groups, we could perform a random walk between investor nodes in the network, and in each step visit a neighboring node proportional to the link weights. In this approach, a group would be a number of investors where the random walker stays for a relatively long time before moving to other investors. However, we cannot identify unambiguous groups simply by performing such dynamics on the network, and therefore we need an extended method. Fortunately, exactly those dynamics are implemented in an existing community detection algorithm, namely the map equation \cite{rosvall2008maps, rosvall2009map}. For network analysis, this algorithm is referred to as Infomap. This algorithm has proven to be one of the most efficient community-detection methods in comparative studies \cite{lancichinetti2009community, aldecoa2013exploring}. In the case of similarity networks, the algorithm identifies groups of investors with strong similarities in portfolio structure, which is precisely what we are looking for.

%\begin{figure}%[!htbp]
%\centering
%\includegraphics[scale=0.37]{Figure-1Bohlin.eps}
%\caption{\textbf{Network reduction technique.} Example of similarity network reduction from single investor level to portfolio level. (A) Network at individual level with nine investors and three different portfolio structures $a$, $b$ and $c$. Investors are connected with links of value 1, $s_{ab}$ and $s_{bc}$. 
%(B) Reduced network at portfolio level, with the three portfolio structures as nodes. Self-links of portfolio nodes represent links between investors with equal portfolio structure. The weight $w_i$ for the self-link of portfolio $i$, with $n_i$ investors, is calculated according to $w_i = n_i(n_i-1)/2$. The link weight $w_{km}$ between portfolios $k$ and $m$, with $n_k$ and $n_m$ investors, respectively, is calculated according to $w_{km} = w_{mk} = n_{k}n_{m}s_{km}$, where $s_{km}$ is the link weight between two investors in each portfolio structure. \label{fig:reduced}}
%\end{figure}

To find a group representation that accounts for both portfolio similarity and market structure, we use the Infomap algorithm with the hierarchical clustering option. This option provides a division of nodes, i.e., portfolio structures, into top-level clusters, consecutively divided into smaller subclusters. We are, however, interested in the division of individual investors, and this division can be obtained simply by mapping each portfolio to its corresponding investors. In this way, we get a categorization of investors into groups that represent similar, but not necessarily identical, portfolio structures. The categorization provides an overview of the stock market, and describes how investors in the Swedish stock market structure their portfolios.

\subsection*{Trading similarity of investors with similar portfolios}
We want to examine if investors with similar portfolio structures tend to trade more similarly than other investors. To determine if two investors are similar, we consider the categorization of investors into groups and study if investors from the same group, i.e., investors with similar portfolio structures, trade in a more similar way than investors outside the group. Trading comparisons for investors of a specific group are made in a bootstrap procedure in the following way: First, we randomly choose a set of investors from the group and compute their aggregated trading vector. Next, we randomly choose two other sets of investors, one set from the same group and one set created from investors that do not belong to the group. We compute the aggregated trading vectors of the two sets and calculate their trading similarity in relation to the first trading vector. In this way, the similarity values make it possible to compare trades within the group to trades outside the group. The detailed procedure looks like:

\begin{itemize}
\itemsep0em
\item[] For each group $G$, repeat $N$ times
	\begin{enumerate}
	\itemsep0em
		\item Randomly choose a set of investors $i_1$ with set size $n$ from $G$. 
		\item Randomly choose another set of investors $i_2$ with set size $n$ from $G$.
		\item Randomly choose a set of investors $i_3$ with set size $n$ outside $G$.
		\item Compute trading vector similarity $s_{inside} = sim(i_1, i_2)$ and $s_{outside} = sim(i_1, i_3)$.
		\item Collect data difference, $\delta = s_{inside} - s_{outside}$, and indicator $I$. $I = 1$, if $\delta > 0$, $I = 0$ otherwise 
	\end{enumerate}
\end{itemize}

In each iteration, we examine whether trades within a group are more similar than trades outside groups, and we repeat this procedure 1000 times. For each group we search for the investor set size that makes the within-group trades significantly more similar than outside trades in the comparison. If portfolio structure and trading were totally dependent, we would have set size 1 for all groups, since this would imply that we can learn about trading of investors in the same group by looking at the trading of only one single investor in the same group. However, the trading data are not very extent for single investors, and therefore we need to compare the aggregated trading of a set of investors to obtain useful information. To measure the trading similarity of a group, we search for the set size that is needed for significance, i.e., the number of investors that are needed so that 95\% of trading comparisons are larger within the group than outside.

% Results and Discussion can be combined.
\section*{Results and Discussion}

\subsection*{Stock portfolio similarity and trading similarity}
Stock portfolios of investors differ by orders of magnitude, both when considering the number of shareholdings and the total value. To unify the portfolio structure, we therefore represent investor holdings in normalized portfolio vectors. This representation considers investment distribution and not the magnitude of investments, which means that two portfolio vectors can be similar even if the total value of the portfolios differs. Analogously, the vector representation is also used to unify the investor trading in trading vectors. When we construct portfolio vectors for the 100,161 investors in the data, we find 52,115 different vectors. Interestingly, only 2,652 portfolio vectors are needed to cover 50\% of all investors, which shows that a large proportion of investors distribute their capital in a similar way. Many investors have capital invested in only one stock, and consequently hold portfolios that are not diversified at all. When we construct the trading vectors, we find that 32,970 vectors are needed to cover all existing trading strategies during the period. The number of trading vectors is smaller than for portfolio vectors because many investors only invest in one or a few stocks.

\begin{figure}[!htbp]
\centering
\includegraphics[scale=1.00 ]{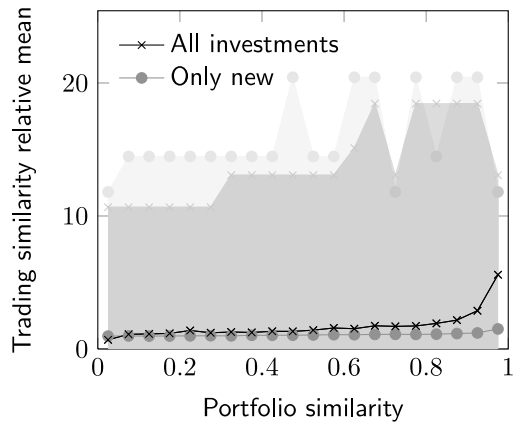}
\caption{\textbf{Relationship between portfolio similarity and trading similarity.} Trading similarity relative to mean value for trading until December 30, 2011, versus portfolio similarity at start June 30, 2009. The figure shows all pairwise comparisons of investors, with portfolio similarity values on the x-axis and trading similarity values on the y-axis. The opaque lines show trading similarity relative to mean value of all trading similarities, as a function of portfolio similarity, and the shaded transparent areas show the non-parametric 95\% confidence intervals. The blue line with circles shows the relationship when all investments of investors are considered, and the red line with x-marks shows the relationship when only investments in new stocks are considered, i.e., investments in stocks that the investors does not already hold. The large variations are results of many zero values for trading similarities, since individual investors do not trade to such a large extent.
 \label{fig:similarity} } 
\renewcommand{\thefigure}{\arabic{figure}}
\end{figure}

Individual investors are more likely to invest in stocks that attract their attention, due to the difficulty of searching among all available stocks. Attention therefore greatly influences individual investor trading decisions \cite{barber2008all}, and the attention-grabbing stocks are naturally the stocks that investors already hold. The combination of the attention bias and the tendency of people to act similarly to their peers, as in, for example, local bias \cite{seasholes2010individual, bodnaruk2009proximity}, gives rise to an interesting question. If individual investors hold portfolios concentrated in only a few stocks, have a preference for investing in stocks they already hold, and also tend to act in accordance with similar investors, does this imply that there is a connection between portfolio structure and trading similarity? The portfolio and trading vectors make it possible to evaluate the question and compare investors, and in Figure \ref{fig:similarity}, we show the relationship between portfolio vector similarity and trading vector similarity. The variation in the data is large, but a trend can be seen in the case when all investments are considered; the more similar the portfolio structure, the more similar the trading. The relationship is evident for portfolio similarity values greater than 0.9, which suggests that these similarity values hold important information. The observed relationship between portfolio structure and trading could be explained with homophily, i.e., the tendency of individuals to engage in similar activities to their peers. This tendency can sometimes make it hard to determine from observational data whether a similarity in behavior exists because two individuals are similar, or because one individual's behavior has influenced the other. Because of the nature of the shareholding data, it is difficult to determine causal reasons for the observed similarities, but since we are primarily interested in the connection between portfolios and trading similarity, this is not an issue.

%\begin{figure}%[!htbp]
%\centering
%\includegraphics[scale=1.00 ]{Figure-2Bohlin.eps}
%\caption{\textbf{Relationship between portfolio similarity and trading similarity.} Trading similarity relative to mean value for trading until December 30, 2011, versus portfolio similarity at start June 30, 2009. The figure shows all pairwise comparisons of investors, with portfolio similarity values on the x-axis and trading similarity values on the y-axis. The opaque lines show trading similarity relative to mean value of all trading similarities, as a function of portfolio similarity, and the shaded transparent areas show the non-parametric 95\% confidence intervals. The blue line with circles shows the relationship when all investments of investors are considered, and the red line with x-marks shows the relationship when only investments in new stocks are considered, i.e., investments in stocks that the investors does not already hold. The large variations are results of many zero values for trading similarities, since individual investors do not trade to such a large extent.
% \label{fig:similarity} } 
%\renewcommand{\thefigure}{\arabic{figure}}
%\end{figure}

\begin{figure}[!htbp]
\centering
\includegraphics[scale=0.60]{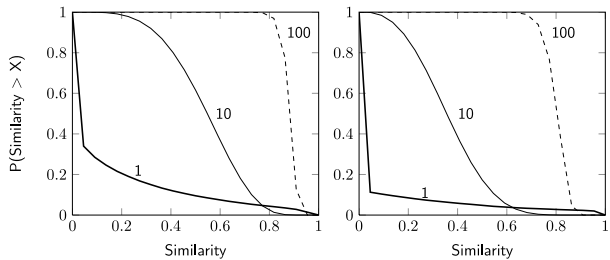}
\caption{\textbf{Portfolio similarity and trading similarity distributions.} Distributions of investor portfolio similarity and trading similarity for groups consisting of 1, 10 and 100 randomly chosen investors. Portfolio similarities are computed for June 30, 2009, and trading similarities are computed from the positive changes in the portfolios until December 30, 2011. The figure shows that a large proportion of investor comparisons are zero in the case of 1-investor comparisons, both in the case of portfolio similarity and trading similarity. For larger groups, the similarity values increase and concentrate to a narrower interval.  \label{fig:cdf}} 
\end{figure}

Comparisons of single investors result in a large proportion of similarity values that become zero, both in the comparisons of portfolio and trading vectors. This means that many investors neither hold nor trade similar stocks, and therefore the evaluation of single investor comparisons becomes problematic. To overcome this problem and be able to compare investors, as a first approach, we created groups of randomly chosen investors, and compared the group's aggregated portfolio and trading vectors to other groups of equal size. The aggregated portfolio and trading vectors are constructed as the mean investment distributions of all investors in the group. The distributions for portfolio and trading similarities, with group sizes 1, 10 and 100, are shown in Figure \ref{fig:cdf}. First of all, the figure illustrates why we want to group investors, since larger groups decrease the number of similarity comparisons that become zero. The figure also shows why we do not want to form these groups randomly, as larger groups cause similarity values to end up in a narrower interval. This shift is a result of the random group formation and demonstrates that the information in such groups is limited, since portfolio dependencies disappear when investors are chosen randomly. Consequently, both group size and how we aggregate groups are important factors when we examine the relationship between portfolio structure and trading,

%\begin{figure}%[!htbp]
%\centering
%\includegraphics[scale=1.00 ]{Figure-3Bohlin.eps}
%\caption{\textbf{Portfolio similarity and trading similarity distributions.} Distributions of investor portfolio similarity and trading similarity for groups consisting of 1, 10 and 100 randomly chosen investors. Portfolio similarities are computed for June 30, 2009, and trading similarities are computed from the positive changes in the portfolios until December 30, 2011. The figure shows that a large proportion of investor comparisons are zero in the case of 1-investor comparisons, both in the case of portfolio similarity and trading similarity. For larger groups, the similarity values increase and concentrate to a narrower interval.  \label{fig:cdf}} 
%\end{figure}

\subsection*{Groups of similar investors from similarity network analysis}
To find groups and analyze the aggregated trading behavior of investors, we model the shareholding data as a network with investors as nodes connected by links according to portfolio similarity. The network approach creates a similarity network, and we analyze this network with the community-detection algorithm Infomap \cite{rosvall2008maps} to identify groups of similar investors. The groups describe how investors in the Swedish stock market structure their portfolios, and the basic properties for the ten largest top-level groups can be seen in Table \ref{tab:stats}. It is worth noticing that more than two-thirds of all investors are included in the ten largest groups. Despite the almost endless number of ways for individual investors to structure their portfolios, the analysis shows that a few related investment strategies are favored. 

The investor groups represent related portfolio structures, and in each group we find some stocks that a large proportion of the investors hold. These top stocks constitute the main connectors between investors in the group. The Ericsson B-stock, which is the stock held by most investors, represents the top stock in the first and largest group. Almost three quarters of the around 25,000 investors in the first group hold shares in Ericsson B. General recommendations on how to invest in the stock market state that diversified portfolios are preferred, but investors still seem to make the choice to hold underdiversified portfolios \cite{goetzmann2005individual}. As a result of this bias, the mean number of stocks held by individual investors is relatively small, which, in turn, makes it possible to identify groups of investors with some specific stock structures in common. When considering portfolio diversity it is worth noticing the possibility that investors also can have capital invested in, for example, diverse funds, but such secondary ownership is not included in the analysis. Individual investors tend to hold only a few different stocks, and this limitation can actually be beneficial, since gathering information on stocks requires resources \cite{merton1987simple}. Individual investors seldom have resources to gather information on more than a few stocks, and informed investors therefore tend to concentrate their portfolios in the stocks in which they hold an informational advantage \cite{ivkovic2008portfolio}.

\subsection*{Similar portfolio structure infers similar trading}
\begin{figure}[!htbp]
\centering
\includegraphics[scale=1.00 ]{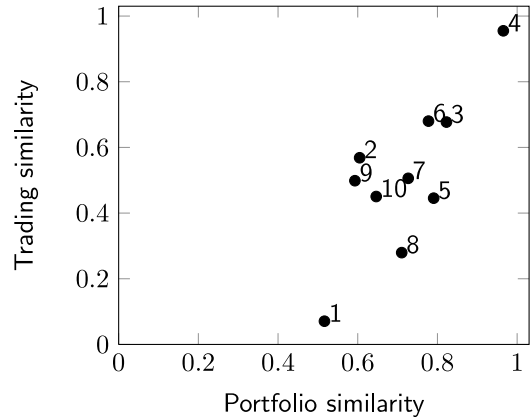}
\caption{\textbf{Group differences in portfolio similarity and trading similarity.} Relation between mean trading similarity and mean portfolio similarity for the investors of the ten largest groups obtained in the analysis of the similarity network. The large and diverse first cluster neither score high on portfolio similarity nor trading similarity, while group four seems to have most homogeneous investors, both when considering portfolio similarity and trading similarity. The portfolio similarity values are all greater than 0.5, since the groups are created with portfolio similarity as a condition. \label{fig:cluster_similarity} }
\end{figure}

The investor groups make it possible to compare the trading of investors with similar portfolio structures. However, the relationship between portfolio structure and trading behavior is dependent on what stocks investors hold, and therefore the relational effect varies between groups. To investigate this relationship, we use a bootstrap procedure in which we compare within-group trades to outside-group trades and search for the investor set size that makes trades significantly more similar within the group. The significant set size specifies the number of investors from the group that is needed for significance in trading similarity, i.e., the number of investors that are needed so that 95\% of the comparisons between aggregated trading vectors are larger within the group than outside. The results are presented in Table \ref{tab:stats}, and we can see that only one investor is required for significance in group 4, while 43 investors are needed in group 1. The number varies between groups, which means that investors with certain portfolio structures tend to trade more similarly than others. The group differences are illustrated in Figure \ref{fig:cluster_similarity}, where the mean trading similarity is shown in relation to mean portfolio similarity, for all investors within the groups. Noticeable is that group 1 has relatively low scores for both portfolio and trading similarity, which can be explained by the fact that the group is large and therefore diverse when it comes to both portfolio structure and trading. Group 4, on the other hand, has a relatively high similarity score for both portfolio and trading similarity. This suggests that the investors within group 4 are more homogeneous than investors in other groups when considering both portfolio structure and trading behavior. Unique to this group is the Saab B-stock, which is held by all investors in the group. It is also interesting to compare the trading behavior of group 2 and 8, since group 2 has a lower portfolio similarity score, but still a higher trading similarity score than group 8. An explanation for these differences could possibly be found by looking at the top stocks of each group, see Table \ref{tab:stats}. The three top stocks of group 2 are all in the car industry sector, while the three top stocks in group 8 are from three different sectors, mining industry, telecommunication and technology. Group 2 therefore seems to represent a more homogeneous ownership, and consequently the investors in the group trade more similarly than the investors in the more diverse group 8.

{\small
\begin{center}
\begin{table}[!htbp]
\caption{\label{tab:stats} Properties of the ten largest groups obtained from clustering the similarity network with 100,161 investors. \textit{Investors} shows the number of investors in the group. \textit{Mean stocks} reports the mean number of portfolio holdings for the investors in the group. \textit{Significant set size} states the set size that is needed for trading significance, i.e., the number of investors that are needed so that 95\% of the trading comparisons are larger within the group than outside. \textit{Top stocks} shows the stocks held by most investors in the group and the corresponding proportion of investors in the group that hold the stock.}
\hfill{}
\renewcommand{\arraystretch}{0.5}% Tighter
\begin{tabular}{c c c c r @{ }l }
\hline
\textbf{Group} & \textbf{Investors} & \textbf{Mean} & \textbf{Significant}  & \textbf{Top} & \textbf{stocks} \\
\textbf{} & \textbf{} & \textbf{stocks} & \textbf{set size}  & \textbf{} & \textbf{} \\
\hline
1 & 25,539 &  2.0 & 43 & $72$\% & Ericsson B\\
&  &&& $44$\% &TeliaSonera\\
&  &&& $8$\% &Volvo B\\ \noalign{\smallskip}
2 & 12,289 & 2.1 & 5 & $81$\%  &Volvo A  \\
&  &&& $36$\% &Volvo B\\
&  &&& $35$\% &Scania A\\ \noalign{\smallskip}
3 & 5,793  & 2.4  & 4  & $98$\% &Sandvik\\
&  &&& $25$\% &Ericsson B\\
&  &&& $15$\% &Seco Tools B\\ \noalign{\smallskip}
4 & 3,769  & 1.2& 1 & $100$\% &Saab B \\
&  &&& $8$\% &TeliaSonera\\
&  &&& $4$\% &Ericsson B\\ \noalign{\smallskip}
5 & 4,589  & 2.5 & 21 & $98$\% &SEB A\\
&  &&& $28$\% &Ericsson B\\
&  &&& $22$\% &TeliaSonera\\ \noalign{\smallskip}
6 & 3,093  & 3.8& 6  & $100$\% &Skanska B \\
&  &&& $35$\% &Ericsson B\\
&  &&& $32$\% &Fabege\\ \noalign{\smallskip}
7 & 3,839   & 3.6& 33  & $100$\% &Nordea\\
&  &&& $44$\% &Ericsson B\\
&  &&& $39$\% &TeliaSonera\\ \noalign{\smallskip}
8 & 2,891   & 3.2 &38 & $100$\% &Boliden \\
&  &&& $43$\% &Ericsson B\\
&  &&& $23$\% &TeliaSonera\\ \noalign{\smallskip}
9 & 3,056 & 5.5 & 20 & $91$\% &Handelsb.~A \\
&  &&& $45$\% &Handelsb.~B\\
&  &&& $42$\% &Ericsson B\\ \noalign{\smallskip}
10 & 2,923  & 4.4 &38 & $100$\% &Investor B \\
&  &&& $49$\% &Ericsson B\\
&  &&& $23$\% &TeliaSonera\\
\mybottomrule
\end{tabular}
\hfill{}
\end{table}
\end{center}
}

%
%\begin{figure}%[!htbp]
%\centering
%\includegraphics[scale=1.00 ]{Figure-4Bohlin.eps}
%\caption{\textbf{Group differences in portfolio similarity and trading similarity.} Relation between mean trading similarity and mean portfolio similarity for the investors of the ten largest groups obtained in the analysis of the similarity network. The large and diverse first cluster neither score high on portfolio similarity nor trading similarity, while group four seems to have most homogeneous investors, both when considering portfolio similarity and trading similarity. The portfolio similarity values are all greater than 0.5, since the groups are created with portfolio similarity as a condition. \label{fig:cluster_similarity} }
%\end{figure}

The group trading similarity may be due to that investors of the same group base trading decisions on similar information because they, for example, possess the same information sources, such as web sites or television \cite{bakshy2012role}. These information sources are more likely to be similar if investors share common interests, which for group 2 potentially could be cars or the automotive industry. There is also evidence of communication among stock market investors, which suggests that investors exchange information about trading in discussions with their peers \cite{shiller1989survey, ivkovic2007information}. Accordingly, social interaction is an influential factor when it comes to stock market trading \cite{hong2004social}. Therefore, our empirical findings on stock portfolio structure could in principle be used to refine multi-agent based order book models with different types of agents \cite{preis2006multi}. However, more research is needed to bridge the gap in time scales between long term investments and short term trades.

To put the results in the context of previous work, we consider some studies that have examined the joint behavior of individual investors, although not from a network perspective. Ref. \cite{barber2009retail} analyzed household trading and found that trading was highly correlated and persistent. The study observed that individual investors tend to react to the same kind of behavioral biases at, or around, the same time. Such behavioral biases could lead to associated trading for related investors, and an explanation for trading similarity could therefore be that similar investors seek and receive similar information over time, and correspondingly trade in accordance. Another explanation to the trading similarity among groups relates to investment herding \cite{grinblatt1995momentum}, where some investors change their portfolio in the same way as a leading group of investors which they trust. In the end, it is interesting to consider the fact that the stock portfolio of an investor actually reflects the aggregated result of all past trades done by the investor, and therefore a relationship between portfolio structure and trading already exists.

\section*{Conclusion}
We show that there is a relationship between stock portfolio structure and trading, namely, that individual investors with similar portfolio structures tend to trade in a similar way. To analyze this relationship, we use real stock market data and a procedure that is threefold. First, we find that comparisons of portfolio and trading similarity for single investors show a large variation, and therefore the data must be analyzed on an aggregated level. Second, we find that the stock market displays a structure among its investors, with groups that represent investors with similar portfolio structures. Third, we find that investors with similar portfolios, to a greater extent, trade in a similar way.

The results show that the stock portfolios of individual investors hold meaningful information, which could be beneficial in the analysis of individual trading behavior. The use of new data sources in economics could improve our understanding of dynamics in financial systems and make it possible to develop models for inferring market reactions. However, even though new and previously unused data can provide important information that relates to market dynamics, the problem of evaluating whether the featured relations are causal or not still persists. Therefore, while future work on the relationship between portfolio and trading includes examining the results from an economical perspective and connecting them to actual market dynamics, the general goal of future work in finance will be to further explore causality when connecting data to dynamics.

%Understanding the relationship between individual investor trading and the market is important for interpreting market dynamics, and examining stock portfolios could help us to better explain the behavior of stock market systems. 

% Do NOT remove this, even if you are not including acknowledgments
\section*{Acknowledgements}
We are grateful to Euroclear Sweden AB for providing the shareholding data on the Swedish stock market, and to Krister Modin, who helped with advice and interpretation of the data. 

%M.R.~was supported by Swedish Research Council grant 2012-3729.

\section*{Author contributions}
LB and MR derived the research project and wrote the manuscript. LB performed numerical simulations.

%\section*{References}
% The bibtex filename
%\bibliographystyle{plos2009}
\bibliographystyle{unsrt}
\bibliography{arXiv_Portfolio_Structure_Infers_Trading-Bohlin_Rosvall}

\end{document}